\documentclass{tufte-handout}

\title{Theory is dead. Long live theory! For a $21^{st}$  century statistical physics of life \thanks{This essay was written for the  CZI Technical Journal (modeled after the Bell Labs Technical Journal) as part of the CZI theory group. The thoughts expressed is this essay are my own. However, they have been shaped by, and greatly benefited from, conversations with the CZI Theory Group (especially Jan\'e Kondev), my collaborators, and the postdocs and students who have come through the group. It goes without saying that the deficiencies and flaws are my own. Given the essay format, I have kept citations to a minimum. Finally, 
I would like to acknowledge support from the Chan-Zuckergberg Initiative.

This piece is motivated by my recent rereading of the history of statistical physics, where philosophical essays and argument were as central to the discussion as technical ones, and a recent historical account of the Theoretical Biology Club in Cambridge ({\it The Life Organic: The Theoretical Biology Club and the Roots of Epigenetics} by Erik L. Peterson. )}  }

\author[slacker physics gang]{Pankaj Mehta}

\usepackage{graphicx} 
  \setkeys{Gin}{width=\linewidth,totalheight=\textheight,keepaspectratio}
  \graphicspath{{graphics/}} 
\usepackage{amsmath}  
\usepackage{booktabs} 
\usepackage{units}    
\usepackage{multicol} 
\usepackage{lipsum}   
\usepackage{fancyvrb} 
  \fvset{fontsize=\normalsize}

\definecolor{pastel_green}{rgb}{0.18,0.65,0.34}

\def\be{\begin{equation}} 
\def\ee{\end{equation}}
\newcommand \bea {\begin{eqnarray}} 
\newcommand \eea {\end{eqnarray}}

\begin{document}

\maketitle

\begin{abstract}
\noindent
The molecular biology revolution of the last 75 years has transformed our view of living systems. Scientific explanations of biological phenomena are now synonymous with the identification of the genes, proteins, and signaling molecules involved. The hegemony of the molecular paradigm has only become more pronounced as new technologies allow us to make measurements at scale. Combining this wealth of data with new ``artificial intelligence'' techniques, which excel at identifying and capturing intricate statistical relationships in large datasets, is viewed as the future of biological science. The successes of AlphaFold and protein large language models (LLMs) at sequence-structure predictions shows the great promise of this approach. In this worldview, there is little room for physics-style theories of life. Capital ``T'' Theory is dead, to be replaced by  elaborate \emph{statistical models} which are perceived to be better suited to describe the living world, characterized by its immense heterogeneity, historicity, variation, and lack of ``universal laws''.  

Here, we challenge this emerging ``common sense'', laying out a roadmap for developing a theoretical understanding of life. We argue that a $21^{st}$ century theoretical biology must be founded on a new type of statistical physics suited to the living world. Rather than merely constructing \emph{statistical models}, a \emph{statistical theory} requires developing ``quantitative abstractions'' for understanding the gene-organism-environment triad\footnote{This phrase is taken from the works of Richard Lewontin (see for example {\it The Triple Helix: Gene, Organism, and Environment}).}. This necessitates overcoming four major challenges that distinguish living matter: (1) living systems are composed of a large number of heterogeneous parts rather than a large number of identical objects; (2) living systems control and manipulate the physical world in a manner that is extremely different from the ways considered in traditional statistical physics; (3) living systems necessarily operate out of equilibrium; (4) living systems are evolved objects with a function, resulting in new types of constraints that must be imposed on probabilistic ensembles.

Much as our current statistical physical frameworks can be applied to the full gamut of physical systems independent of particular details (e.g. quantum gases, neutron stars, membranes, etc.), the new statistical mechanics of life should be deployable to the full diversity of biological phenomena (e.g. eco-evolutionary processes, metabolic networks, neural systems, etc.). The foundations for such a theory must be build on a few key themes: typicality, localized biological control, a linear response of complex systems with non-reciprocal interactions, biological resource allocation, and learning and adaptation in overparameterized systems. We highlight recent works in these directions and argue that these foundations can be extended to develop a statistical physics of life.

\end{abstract}


\section{Introduction}

A biologists transported from the beginning of the twentieth century to the present day would scarcely recognize their discipline. During the first part of the twentieth century, biology was still natural history, concerned primarily with observing and characterizing the living world. Though evolutionary theory was in ascendancy, vitalism, the idea of that living organisms possessed a special force, was still a widely accepted as an explanation for the unique properties exhibited by living matter. 

The first half of the twentieth century also saw the transformation of many sub-disciplines of biology into an experimental lab science. Embryological experiments highlighted the complex and pliable processes underlying the emergence of form and function in organisms. These experiments ultimately led to demise of vitalism and an embrace of abstract ideas to interpret the robust self-organization observed in embryonic development . Concepts such as induction, organizers, gradients, and fields were a staple of the language used by ``organicists'' to describe and understand development \footnote{see  Nick Hopwood  {\it The Cambridge History of Science} , Chapter 16, Embryology, pp. 285 - 315, 2009 and Donna Haraway, {\it Crystals, Fabrics, and Fields: Metaphors That Shape Embryos, Yale University Press, 1976.} }. If and how these abstractions related to the molecular and biochemical remained a major area of scientific and philosophical debate.

The discovery of DNA and its structural characterization marked the beginning of a fundamental shift in the epistemological underpinnings of the life sciences. Genes and molecules became the currency of biological understanding. Seventy years after the work of Crick, Watson, and Franklin, the molecular paradigm is so hegemonic that when biologists speak of understanding an aspect of life, it is synonymous with identifying the genes and molecules involved. 

A major thrust of recent research has been to move beyond identification of parts and  characterize the interactions between molecules and genes. This has been complemented by a concerted attempt by some scientists to make biology into a quantitative science, ``to move beyond the cartoons'' \footnote{Phillips, Rob, and Stephen R. Quake. "The biological frontier of physics." Physics Today 59.5 (2006): 38-43.} by ``assigning numbers to the arrows''. \footnote{ Ronen, Michal, et al. ``Assigning numbers to the arrows: parameterizing a gene regulation network by using accurate expression kinetics.'' Proceedings of the national academy of sciences 99.16 (2002): 10555-10560} Impressive new technologies, many developed by physicists and engineers, have allowed for precise quantification of  gene and protein expression, interaction strengths, and even temporal dynamics of signaling networks. These observations have been codified into impressive predictive mathematical models, confirming the idea that like physical systems, living systems can be quantitatively described using mathematics.

\section{From high-throughput biology to statistical models}

Over the last two decades, the quantification of biology has been transformed by the advent of high-throughput techniques for measuring and perturbing living systems. The cost of sequencing has dropped a million fold in the last two decades, resulting in an explosion of biological sequence data.\footnote{https://www.genome.gov/about-genomics/fact-sheets/DNA-Sequencing-Costs-Data} The hegemony of the molecule paradigm has only become more pronounced as these new technologies allow us to make measurements at scale. Scientific decision makers  have launched a plethora of projects that seek to comprehensively catalogues living systems (largely in terms of gene expression but also increasingly metabolically and at the level of proteins).  Biological ``atlases''  now exist for every organ in mouse and humans, microbial communities, as well as many diseases including cancers.\footnote{The atlas metaphor as a name is quite interesting since the high-dimensionality of the data often makes it difficult to infer relationships between points, something that is the foundational to the concept of an atlas.}

The sheer quantity of data in these atlases naturally raises the question of how one can meaningfully extract biological insights from these datasets.  This requires grappling with some major technical and conceptual challenges. On the technical side, high-throughput experiments such as single-cell RNA sequencing tend to be  high-dimensional,\footnote{By high-dimensional, I will mean measuring many correlated features at once.} extremely noisy, and exhibit large batch effects.\footnote{Batch effects is commonly used language to denote experiment or lab-specific systematic errors} These properties make it challenging to disentangle technical variations from natural variations. Furthermore, despite the vast experimental scale of current atlases, there remain many aspects of the biology that remain unmeasured. For example, current atlases mostly lack  biochemical measurements of post-translational modifications involved in major signaling pathways.\footnote{The underlying reason for this is that such measurments cannot be easily performed using DNA sequencing or fluorescent proteins. Instead, this requires the manufacturing of specialized antibodies.} 

The task of interpreting large-scale datasets is also hampered by two central conceptual challenges: the role of environment and the challenge of context.  A fundamental property of living systems is that they both shape and respond to the environments in which they grow. While this is most evident in ecology,\footnote{ We have extensively pursued this line of thinking in the context of the important role played by metabolism and cross-feeding in microbial ecology. See for example Goldford, Joshua E., et al. "Emergent simplicity in microbial community assembly." {\it Science} 361.6401 (2018): 469-474, Marsland III, Robert, et al. "Available energy fluxes drive a transition in the diversity, stability, and functional structure of microbial communities." {\it PLoS computational biology} 15.2 (2019): e1006793., Marsland III, Robert, Wenping Cui, and Pankaj Mehta. "A minimal model for microbial biodiversity can reproduce experimentally observed ecological patterns." {\it Scientific reports} 10.1 (2020): 3308.} it is equally true in other areas of biology. For example, in the process of morphogenesis and cellular differentiation, cells actively modify their chemical and mechanical environments through signaling, inexorably intertwining genes, molecules, and environments. Yet, most datasets make no attempt to measure, let alone comprehensively characterize, environmental variables and feedbacks. 

The second major conceptual challenge that must be overcome is the problem of biological context. Much of the complexity of biology can be traced to the fact that components in living systems behave extremely differently depending on the context in which they are found. Just as a word can mean very different things depending on the sentence, \footnote{Consider for example the meaning of blue in the sentences ``The sky was blue.'' and ``The man was blue.''} biological parts often exhibit extreme context dependence. A prominent and practically important example of this is in protein folding where the same amino acid behaves extremely differently depending on what other amino acids it is surrounded by. Similar observations hold at the level of ecosystems, where the behavior of organisms is deeply influenced by what other species and resources are in the environment. 

Faced with these challenges, the community has embraced complex statistical models and dimensional reduction techniques for extracting biological insights. This process has been 
accelerated by the development of large scale models associated with the modern Deep Learning revolution which excel at identifying and extracting patterns from large datasets.\footnote{ For a comprehensive pedagogical review aimed at physicists, please see our article Mehta, Pankaj, et al. "A high-bias, low-variance introduction to machine learning for physicists." {\it Physics reports} 810 (2019): 1-124.} Combining these newly generated  datasets with modern deep learning techniques is viewed as the future of biological science.

The most dramatic example of the immense potential of this research program are the impressive successes of AlphaFold at sequence-structure predictions,\footnote{Jumper, John, et al. "Highly accurate protein structure prediction with AlphaFold." {\it Nature} 596.7873 (2021): 583-589.} and the ability of protein language models to learn generative models for  diverse protein families.\footnote{Madani, Ali, et al. "Large language models generate functional protein sequences across diverse families." {\it Nature Biotechnology} 41.8 (2023): 1099-1106.} \footnote{These techniques build upon earlier works by physicists and system biologists showing the statistics of amino acid conservation within a protein family are sufficient to predict protein properties. The first definitive example of this I know are from the works of Rama Ranganathan Lockless, Steve W., and Rama Ranganathan. "Evolutionarily conserved pathways of energetic connectivity in protein families." Science 286.5438 (1999): 295-299. Also see Halabi, Najeeb, et al. "Protein sectors: evolutionary units of three-dimensional structure." Cell 138.4 (2009): 774-786 and Morcos, Faruck, et al. "Direct-coupling analysis of residue coevolution captures native contacts across many protein families." Proceedings of the National Academy of Sciences 108.49 (2011): E1293-E1301.} By leveraging vast public protein sequence and structure repositories (for example  Progen, a prominent protein language model, was trained on 280 millions sequences from over 19,000 Pfam families), these models essentially allow for the generation and structural prediction of novel sequences. \footnote{ A major caveat is that this is really only successful for novel sequence that encode for proteins similar to those found in the training data.} These results suggest that deep learning techniques may be equally adept at capturing the intricacies of biological context as linguistic context.\footnote{ This is especially true of self-supervised learning techniques and attention-based transformer architectures. The Reinforcement Learning underlying Deep Mind's models seems to be less transferable.}

\section{Foundation models and the death of Theory}

These recent developments have reinforced the long-held sentiment that perhaps biological systems are just too complex to be amenable to physics-style approaches.  Within this emerging ``common sense'', the immense  heterogeneity, historicity, and the strong role played by chance and variation prevent biology from being tackled with techniques that have been so successfully deployed to describe the physical world. In this regard, it is interesting to quote the biologist and philosopher of science Ernst Mayr who insists that in contrast to physics ``Most theories in biology are based not on laws but on concepts.'' \footnote{Ernst Mayer {\it What Makes Biology Unique?} p 43.} In Mayr's view, the underlying reason for this is that living organisms are defined by dual causation. ``In contrast to purely physical processes, these biological ones are controlled not only by natural laws but also by genetic programs.'' \footnote{Ernst Mayer {\it What Makes Biology Unique?} p 30.}

The views expressed by Mayr are implicitly held by the vast majority of biologists today. The success of recent Deep Learning approaches have reinforced these ideas. The prevailing sentiment today can be summarized as follows:  Capital ``T'' Theory is dead. We should instead focus  building elaborate \emph{statistical models} which are perceived to be better suited to the task of describing living systems.\footnote{There is even catchy new buzzword for this idea, foundation models.} The role of biological thinking in this approach is simply to curate good data, conduct experiments, and supply enough ``inductive biases'' so that foundation models can identify patterns and distill biological knowledge. The most extreme version of this vision imagines an automated feedback loop between statistical models, predictions, and new experiments, with artisanal science playing a steadily decreasing role. The role of the scientists is simply to deploy the insights encoded in foundation models to engineer new tools and therapies. Theory and abstraction are prominently absent from this version of the scientific future.

\section{Challenging the emerging ``common sense''}

Vast amounts of resources are being mobilized in pursuit of this scientific vision. Yet, there remain deep philosophical and technical reasons to be skeptical that elaborate statical models can play the same role in biology that theory plays in physics. There is absolutely no doubt that modern machine learning techniques have become, and will continue to be, a major computational tool for analyzing high-dimensional biological data. After all they excel at wide variety of tasks that are central to building a theory of biology:  identifying patterns,  finding low-dimensional structure in high-dimensional data, learning probabilistic models, and constructing simple mathematical representations of complex probabilistic structures.\footnote{One of the most interesting developments is the ability to use self-supervised learning objectives to embed almost anything into a vector space. Prominent examples include things like transformers and CLIP. (see Radford, Alec, et al. "Learning transferable visual models from natural language supervision." International conference on machine learning. PMLR, 2021.)} But it is much less clear how such statistical models can replace theory. After all theories, through abstraction, allow for the generalization of insights well beyond the settings in which they were first gleaned.\footnote{It is hard to understand how modern statistical learning techniques could every produce Newton's theory of gravity or Einstein's relativity. That being said, given enough data one can imagine them discovering the analogue of Kepler's laws -- no small feat.}

 Over-parameterized statistical models, including the most advanced examples such as Large Language Models, are fundamentally about interpolation rather than extrapolation.\footnote{ Our group has spent a lot of time trying to develop intuition for why statistics still works when there are many more parameters than datapoints. For an in depth but accessible discussion, see Rocks, Jason W., and Pankaj Mehta. "Memorizing without overfitting: Bias, variance, and interpolation in overparameterized models." Physical review research 4.1 (2022): 013201.} They exploit the considerable structure found in natural datasets to interpolate high-dimensional data distributions and learn probabilistic mappings between high-dimensional spaces. What these models aren't designed to do is precisely what theory excels at: extrapolation to new settings, innovation, and abstraction. This is especially true in the face of limited amounts of data which is the norm in science. 
 
Modern statistical models seem better suited at automating expensive, repetitive tasks such as sequence-structure prediction. As incredible as it is,  AlphaFold, at its essence, is simply a differentiable map between the spaces of sequences and structures in the PDB.  One should not be surprised that AlphaFold fails precisely at situations that are poorly represented in current protein databases: multimeric complexes, intrinsically disordered proteins, conformational ensembles, allosteric changes, protein dynamics, and mutational tasks.

For this reason, there is good reason to believe that the mathematical representations learned by such statistical models are a starting point for understanding, rather than an end itself. What is required is precisely what  statistical models cannot provide: ``quantitative abstractions'' for understanding the gene-organism-environment triad.\footnote{The Silicon Valley, ''Artificial General Intelligence (AGI) is imminent'' crowd will strongly disagree with the sentiments of these last paragraphs. But I have seen very little evidence to suggest otherwise. Many Deep Learning researchers share my skepticism, though they course do not get same the media headlines as AGI boosters. A notable exception is Yann Lecun. (see ``Meta's AI Chief Yann Lecun on AGI, Open-Source, and AI Risk.'' {\it Time} 2024.)} 
 
\section{Why must the statistical physics of life be different?}

It is clear we need a $21^{st}$ century theoretical biology. The foundations of a such a theory must be rooted in a new type of statistical mechanics tailored to the living world. The pillars of twentieth century theoretical physics -- symmetry, energy functionals, and optimization -- are ill-suited to be the core mathematical techniques for describing life.\footnote{Of course, there are notable exceptions. Symmetry remains important when thinking about certain developmental processes and the emergence of form. There are physicists and biologists -- the adaptationist school -- that view evolution as an optimization engine. However, I think even if this is true, what is really important are the plethora of constraints placed by biology on this process. The constraints are likely so numerous and diverse in nature, that it is unclear if this line of thinking can yield much insight beyond some well chosen examples where these constraints are easy to identify.} 

 Living systems pose new challenges largely absent in the inanimate world:

\begin{enumerate}

\item{Biological systems are composed of a large number of heterogeneous parts rather than a large number of indistinguishable objects. Statistical physics theories, whether classical or quantum, are largely designed to handle large number of identical degrees of freedoms, be it particles in a gas, spins in a magnet, or vortices in a superfluid. In contrast, in living system, even presumably identical degrees of freedom can change their properties. For example, cells in a monoculture can, and often do, behave very differently from each other. This is because as discussed above, living systems are subject to dual causation. Their behavior is governed not only by physical processes but also by genetic programs designed to allow them to adapt and change in response to their local environments.}

\item{Another challenge posed by living systems pertains to the nature of biological control. Statistical mechanics is at its core is a theory of control. The difference between work and heat (and the origin of entropy) is precisely the presence of degrees of freedom that cannot be controlled with infinite precision. This observation is at the root of Loschmidt's famous objection to Boltzmann's H-theorem (now know as Loschmidt's paradox).\footnote{Loschmidt imagined a thought experiment where we had control over every particle. He then imagined letting the particles evolve in time, and at some later time point, reversing all the velocities of the particles. Since Newton's laws are time-reversal invariant, the system then ends up back at the initial state and entropy does not increase. This argument clearly requires the ability to measure and manipulate the velocities of every degree-of-freedom with infinite precision.} As currently formulated, statistical mechanics is almost entirely about \emph{external}  control of an \emph{extensive} number of degrees of freedom. These theories are designed to describe systems where an experimentalist can manipulate generalized thermodynamic forces or fields. In contrast, in living systems, the control variables are often local, autonomous, and dynamic.\footnote{These thoughts were developed in close collaboration with Jan\'e Kondev.} Rather than manipulate global phases, cells often locally nucleate a phase (consider the central role played by nucleators in diverse biological processes ranging from phase-separation to the formation of local actin structures such as spindles) and dynamically change control variables in response to their environment (for example, feedbacks that change surface tensions and line tensions of cells in tissues in response to mechanical cues). This is in stark contrast to the static global control variables coupled to extensive quantities considered in classical statistical physics. }

\item{A defining feature of living systems is that they are fundamentally nonequilibrium -- they ``feed on negative entropy'' to use Schr\"ondinger's colorful phrase.\footnote{Ewrin Schr\"ondinger. {\it What is life?}} Living systems operate out-of-equilibrium at all scales: from the nanometer size steps of DNA polymerase, to sub-cellular structures, to tissues, organisms, and ecosystems. Yet, our most successful statistical mechanics theories and formalism are designed almost entirely for equilibrium systems.}

\item{Living systems are also evolved objects with function, namely ``there is a small subset of properties that is of great importance to biology, and that evolutionary choices have shaped biological systems so that they function well.''\footnote{This definition is taken from John Hopfield's insightful essay ``Now What?'' https://pni.princeton.edu/document/1136} How to write a statistical mechanics of matter with function raises a number of pressing questions. How do we include functional and evolutionary constraints when constructing probabilistic ensembles of microstates? How and when do we include control variables that couple to these constraints?\footnote{This statement is inspired by thinking about generalized thermodynamic forces as objects that enforce constraint on averages as in the Jaynes formulation of statistical physics.} We must also understand how living systems can evolve such functions in the first place. Any credible statistical mechanics of living systems needs to grapple with evolution, including taking seriously the difficulties of mapping genotypes to phenotype and the fundamental role played by the environment in this process.\footnote{I would say the contributions of physicists to developing a statistical mechanics approach to population genetics has been one of the major achievements of the physics of living systems community in the last twenty five years. That being said, as in all of population genetics, much of what makes biology so rich and interesting is prominently absent from these theories.} }
\end{enumerate}

\section{What is to be done?}

Much as statistical physics can be applied to the full gamut of physical systems independent of particular details (e.g. quantum gases, neutron stars, fluids, membranes, phase transitions, etc.), the new statistical mechanics of life should be deployable to the full diversity of biological phenomena (e.g. eco-evolutionary processes, tissue mechanics, metabolic networks, self-organization of cellular structures, behavior, neural systems, etc.).\footnote{For a small example, of how diverse systems can be treated with similar formalism please see my accompanying CZI paper.} Developing such a theory represents the holy-grail of those committed to the idea that the complexity of life does not preclude the idea that we can understand it with the same level of rigor as we do the physical world. So how do we proceed?  Here, I outline some key concepts that I believe can serve as the foundations for such a theory.

\begin{enumerate}

\item{\emph{Typicality with constraints}. One of the most interesting ideas to emerge in statistical physics in the second half of the twentieth century was the idea of typicality. Wigner showed that the statistical structure of nuclear resonances in the Uranium atom could be reproduced by modeling the Hamiltonian as a large symmetric random matrix.\footnote{Wigner, Eugene P. "Random matrices in physics." SIAM review 9.1 (1967): 1-23.}  Winger's conjectured that when a system becomes complex enough, many of its statistical properties will be ``typical'' and indistinguishable from a large random system subject to the same physical constraints as the object of study (i.e. is drawn from the same ensemble). In the case of Uranium, the relevant physical constraint Wigner identified was unitarity, which is why he considered \emph{symmetric} random matrices to model the Hamiltonian of Uranium. More recently, this line of thinking has found great success in condensed matter theory in the context of Eigenstate Thermalization Hypothesis, which uses typicality to explain why even a single eigenstate of a sufficiently large many-body quantum system can be accurately described by statistical mechanics.\footnote{See this very nice review D'Alessio, Luca, et al. "From quantum chaos and eigenstate thermalization to statistical mechanics and thermodynamics." {\it Advances in Physics} 65.3 (2016): 239-362.} 

As Freeman Dyson explained, this is a fundamentally  ``new kind of statistical mechanics, in which we renounce exact knowledge not of the state of a system, but the nature of the system itself.''\footnote{see p2 Wigner, Eugene P. "Random matrices in physics." {\it SIAM review} 9.1 (1967): 1-23} What is nice about this approach is that it suggests a natural framework for thinking about living systems. Because it is agnostic to the origin of the constraints, it does not matter whether the relevant constraints originate from physical conservation laws, dynamics, geometry, energetics, evolutionary history, or even from specialized biological functions. For this reason, these ideas offer a promising avenue for thinking about how to model complex biological systems.

In the last few years this program has been successfully applied to understand many properties of complex microbial communities, suggesting that it can be operationalized to gain biological insights \footnote{For 
example, see Goldford, Joshua E., et al. {\it Science} 361.6401 (2018): 469-474., Marsland III, Robert, Wenping Cui, and Pankaj Mehta. {\it Scientific reports} 10.1 (2020): 3308., Ho, Po-Yi, Benjamin H. Good, and Kerwyn Casey Huang. "Competition for fluctuating resources reproduces statistics of species abundance over time across wide-ranging microbiotas." {\it Elife} 11 (2022): e75168.}  Moreover, modern generative models of proteins also naturally fall into this rubric. When stripped of all their bells and whistles, generative models of proteins are probabilistic models for sequence generation that respect the evolutionary and functional constrains embodied in amino acid conservation patterns.\footnote{This was a common perspective on generative protein models until the last five years. Recently, this narrative frame has retreated from popular conscience and been replaced by an almost exclusive focus on the machine learning architectures that enable one to learn such a distribution well.} }

\item{ \emph{Localized biological control}. A much less well developed but equally promising direction is to think about the nature of biological control. Traditionally, in statistical mechanics the control variables are generalized thermodynamics forces conjugate to extensive state variables. However, as discussed above, in biology the control variables are local and dynamic and, as such, have the ability to control degrees of freedom at a much finer level.  This level of increased control is only possible because living systems operate out of equilibrium. Often, even individual degrees of freedom consume energy. For this reason, we need to develop an understanding of how consuming energy allows this increased level of control and what kind of properties are possible because of this.

Such ideas are just starting to be explored in the realm of active matter, stochastic thermodynamics, and the soft-matter community studying ``physical learning''. However, much of this work is still focused on biologically-inspired physical systems rather than living systems themselves. For this reason, it will be important to start trying to formulate and apply these concepts in actual biological systems. This requires working in close collaboration with experiments.\footnote{See Jan\'e Kondev's contribution to the CZI forum.}  Inevitably, a true understanding of biological control will involve synthesizing and expanding ideas from numerous subfields such as nonequilibrium statistical mechanics, stochastic processes, and dynamical systems. If guided closely by experiment, this will allow for the identification of universal features of biological control shared across living matter.}

\item{ \emph{Linear response of complex systems with non-reciprocal interactions}

Another immediate consequence of the nonequilibrium nature of biological systems is that they often interact non-reciprocally -- namely how A affects B is not the same as how B affects A. This is in stark contrast from equilibrium systems where reciprocal interactions are the norm (the most famous example being Newton's third law). This absence of reciprocal interactions is closely related to the fact that forces no longer can be written as gradients of a potential function.  This has major mathematical implications for developing a statistical physics of life, and in particular a theory of linear response. 

Linear response is one of the bedrocks of physics and asks how a system responds to a small change in external parameters. The relationship between the perturbation and the response is encoded in susceptibility matrices. Susceptibility matrices play a central role in all areas of physics ranging from electromagnetism to quantum field theory. Susceptibilities can be used to find effective parameters, identify collective modes, construct perturbation theories, and even design coarse graining schemes.\footnote{The most famous of these is of course the renormalization group.} In equilibrium, susceptibility matrices are symmetric (since interactions are reciprocal) and the left and right modes are identical. However, out of equilibrium, this is no longer true since interactions are nonreciprocal. This leads to a completely different mathematical structure that is poorly understood. 

The physics community has just started investigating linear response of nonreciprocal systems. The place where this is best understood is in the context elasticity and fluid dynamics.\footnote{Fruchart, Michel, Colin Scheibner, and Vincenzo Vitelli. "Odd viscosity and odd elasticity." Annual Review of Condensed Matter Physics 14 (2023): 471-510.} However, even in this setting, much remains unknown. In addition, much more work is needed to understand systems that cannot be easily modeled using differentiable variables or using hydrodynamic arguments where symmetry considerations play a prominent role. This later situation is the norm for most biological systems such as biochemical and signaling networks, eco-evolutionary dynamics,  and neural systems. In such systems, the degrees of freedom are often discrete, heterogenous, and can even disappear from the system.\footnote{We have recently explored these ideas in a recent preprint Goyal, Akshit, Jason W. Rocks, and Pankaj Mehta. "A universal niche geometry governs the response of ecosystems to environmental perturbations." bioRxiv (2024): 2024-03.} Such systems must also often satisfy additional constraints stemming from the biology. How such constraints modify linear response is something about which very little is understood. }

\item{ \emph{Biological resource allocation} Living systems grow and self-replicate. This has implications for all scales of biology, from the sub-cellular to the planetary. Growth requires procuring resources from the environment and allocating these resources across the multitude of processes that sustain life. The most prominent and obvious example of resource allocation is metabolism. But processes as diverse as the nucleation of actin structure in yeast, where nucleators and actin molecules must be divided across multiple actin structures, to ecosystems, where species self-organize to partition resources in the environment, can also be viewed through the lens of resource allocation.

Numerous recent work suggest that despite the complexity of these processes it may be possible to build simple models that capture the essential logic of biological resource allocation. The key theme emerging from these works is that resource allocation is often controlled by a few key bottlenecks and that these bottlenecks serve as natural targets for evolution since the system is particularly sensitive to changes at these locations.\footnote{A beautiful example of this is the recent work from Terry Hwa and collaborators on cellular growth laws reviewed in  Klumpp, Stefan, and Terence Hwa. "Bacterial growth: global effects on gene expression, growth feedback and proteome partition." \it{Current opinion in biotechnology} 28 (2014): 96-102.} Another promising line of works draws on thinking about resource allocation to understand allometry and organism size. It will be interesting to understand if this kind of thinking, which has been so successful in the context of organismal biology and ecology, can also be extended to the sub-cellular scale. 

More generally, we must incorporate the constraints placed by growth and resource allocation into any statistical mechanics of life. How to do this remains one of the major open problems facing our community.
}

\item{ \emph{Learning and adaptation in over-parameterized systems}.\footnote{This line of reasoning will be discussed extensively in an upcoming review for the {\it Annual Review of Biophysics}.} One of the most striking things about living systems for any physicists who enters biology is just how many different components they possess. For example, the biochemical networks that allow cells to respond to their environments often look like a dense, loosely structured collections of interacting components, where the relationship between form and function remains obscure.\footnote{Of course, there are notable exceptions including the bacterial chemotaxis network and photon detection by the early retina.} A similar picture emerges once one starts looking at neurons and their connections, or the interactions of immune cells that mediate immune responses. These objects seem extremely far removed from the modular, streamlined structures that characterize most human engineered objects. These information processing networks seem even more mysterious because there are often numerous (partially) interchangeable components, and removing components often does not degrade performance. Such redundancy and robustness is often taken to be a defining feature of biochemical signaling networks.\footnote{In the systems biology literature, this was often discussed in the context of the lack of ``fine tuning'', but seems to have largely fallen out of favor recently.} and neural systems.\footnote{A beautiful example of this is the stunning work from Eve Marder and collaborators on neurons in central pattern generators. See this recent review Goaillard, Jean-Marc, and Eve Marder. {\it Annual review of neuroscience} 44 (2021): 335-357.}

One promising avenue for making sense of these observations is to draw upon recent insights regarding learning in overparameterized networks. The last few years has seen a major paradigm shift in our understanding of statistics and learning. Until recently, it was commonly believed that training statistical models with many more parameters than training data points would inevitably give rise to overfitting and poor performance. It has become clear that, even for simple models, this is generally not the case.\footnote{This ideas was rediscovered and brought into the popular conscience by a series of works by Belkin and others. Belkin, Mikhail, et al. {\it Proceedings of the National Academy of Sciences} 116.32 (2019): 15849-15854. For an intuitive explanation of this phenomenon rooted in statistical physics see Rocks, Jason W., and Pankaj Mehta. {\it Physical review research} 4.1 (2022): 013201.} Remarkably, in many cases, the more parameters that models have the better they seem to perform (think of Large Language Models).\footnote{This observation has given rise to whole sub-field in ML centered around scaling laws.} And just like biological models, empirical work shows that it is possible to prune modern neural networks, sometimes removing up to 99 percent of parameters without a significant degradation of performance.\footnote{Frankle, Jonathan, and Michael Carbin. "The Lottery Ticket Hypothesis: Finding Sparse, Trainable Neural Networks." International Conference on Learning Representations. 2018.} However, if tries to directly learn using these pruned networks, they generally do not perform well unless one fine tunes their initial parameters close to their initial values in the full network. This suggests that not only does having many parameters not hurt, it may actually make learning complex functions easier than it is in small networks. Finally, just like in biological networks, it is incredibly difficult to construct a simple mapping between parameters and function. Instead, the function seems to be a collective property of the whole network.

There are hints that overparameterization --having many more degrees of freedom than constraints that have to be satisfied -- is also advantageous for learning in biology and biology-inspired settings.\footnote{Note, that in the context of statistical fitting, we can think of each datapoint in the training data set as a constraint on the parameters.} Recent works, suggest that properties such as allostery can be easily learned as long as the underlying mechanical networks are overparameterized.\footnote{See for example Rocks, Jason W. et. al. {\it Proc. Natl. Acad. Sci.}. 2017 } The underlying reason for this is that when the number of degrees of freedom is much larger than the number of constraints one has to satisfy, one can easily find solutions that do not require fine tuning. This has obvious implications for understanding the evolution and complexity of living organisms. All these observations collectively suggest that a better understanding of overparameterization will be a fundamental part of developing a statistical physics of learning and adaptation in living systems. }
 
\vspace{0.2in}
While the topics outlined above layout some key themes that I think will be central to developing a statistical mechanics of life, in no way should they be taken as a comprehensive list. The history of science teaches us that theories often develop in unexpected ways, governed by an invisible internal logic that is hard to predict. Nonetheless, from my current vantage point, these general directions look particularly promising for starting to tame the complexity of biology and place it on a theoretical foundation on par with the physical sciences. 
 
\section{Conclusion}

Twenty first century biology is defined by data. Experimental and computational advances are making it easier to make measurements at scale but our ability to make sense of this data deluge is lagging. This situation was summarized by Noble prize winning biologist Sydney Brenner nearly a decade ago in his retrospective for Alan Turing:
\begin{quote}Technology gives us the tools to analyse organisms at all scales, but
we are drowning in a sea of data and thirsting for some theoretical framework with
which to understand it. Although many believe that `more is better', history tells us
that `least is best'. We need theory and a firm grasp on the nature of the objects we study
to predict the rest.''\footnote{Sydney Brenner. {\it Nature} 482, p.461 (2012)}
\end{quote}
Despite Brenner's warning, we are currently witnessing a turn away from theory and abstraction. Understanding biology has become synonymous with identifying parts and interactions. 

It is widely recognized that this situation is untenable. Biology is in need of new techniques that allow for synthesis. A significant fraction of the scientific community has abandoned traditional theory, placing their hopes in elaborate statistical models, especially those associated with the Deep Learning revolution. While there is no doubt such models will become increasingly important in biology, they are fundamentally different from theory. Theories can generalize insights well beyond the settings in which they were first conceived. Theory has the ability to innovate and unify disparate knowledge into a holistic synthesis. This is in contrast with statistical models which are limited to interpolating the datasets on which they are trained. For this reason, what is urgently needed is a twenty first century statistical physics of life.

But developing a statistical physics of life is full of challenges. Symmetry and optimization, the pillars of twentieth century physics, can no longer serve as the foundations for such a theory. We must turn to new ideas and techniques that can tame the complexity of biology. Doing so will requite moving beyond both twentieth century physics and elaborate statistical models. Perhaps, eighty years after the publication of Shr{\"o}ndiger's \emph{What is Life}, we are finally in a position to ``find a new type of physical law prevailing in [life].''
\end{enumerate}

\end{document}